\documentclass [12pt]{article}
\usepackage{amsmath,amssymb,cite}
\usepackage{appendix}
\usepackage{feynmp}
\usepackage{float}
\usepackage{subfigure}
\usepackage[vcentermath,enableskew]{youngtab}
\usepackage{tabularx}
\usepackage[english]{babel}
\usepackage{graphicx}
\usepackage{indentfirst}
\usepackage{epsfig}
\usepackage{color}
\usepackage{epstopdf}
\usepackage[]{hyperref}
\usepackage[section]{placeins}
\usepackage[stable]{footmisc}
\usepackage{appendix}
\usepackage{tabularx}
\usepackage[english]{babel}
\usepackage{graphicx}
\usepackage{indentfirst}
\usepackage{epsfig}
\usepackage{slashed}
\usepackage{fancyhdr}
\usepackage{quoting}
\usepackage{booktabs,longtable}
\quotingsetup{vskip=0pt}
\numberwithin{equation}{section}

\fancypagestyle{plain}{\fancyhead{}
}

\begin{document}

\title{\textbf{Quantum Perspectivism vs Nietzschean Perspectivism}}

\author{Badis Ydri\\
\textit{Annaba University, Algeria}\\
}


\maketitle
\begin{abstract}
  This is a work of hard physical philosophy, where Quantum Perspectivism is shown to function as both an interpretation of quantum mechanics and a physical model for understanding Nietzsche’s perspectivism. This framework combines quantum logic, the principle of complementarity, and contextuality to examine how perspectives construct reality. In this model, measurements correspond to Perspectives and Meta-Perspectives, represented as Boolean subalgebras and Hilbert sub-lattices within the Hilbert lattice, respectively. The Hilbert lattice itself is reinterpreted as Jung’s Unus Mundus, a unified ontological reality. A metaphysical observation, made by a metaphysical observer, of a given system (World) is identified with the set of all corresponding meta-perspectives in the Hilbert lattice/Unus Mundus, the ocean of reality.

Perspectives, likened to islands in this ocean, correspond to single measurements of a system, capturing the logical structure of observed properties. Meta-perspectives, analogous to continents, represent the synthesis of multiple measurements, providing a broader yet inherently incomplete understanding of the system. This structure emphasizes the complementarity and contextual dependencies of measurements while exposing the limitations of classical objectivity in the quantum domain. Advocating for a perspectival view of both the world and truth, Quantum Perspectivism unites quantum mechanics and Nietzschean philosophy into a cohesive framework for exploring the interplay between consciousness, observation, and reality.

  \end{abstract}



\tableofcontents

\section{Introductory Remarks}
\subsection{Prologue}

Philosophy has long addressed the profound questions of existence, offering conceptual frameworks for understanding the human condition. Theoretical physics, on the other hand, provides a rigorous framework for uncovering the deepest structures of reality. At their intersection lies a fertile ground for inquiry—one that seeks to balance the reflective depth of philosophy with the empirical rigor of physics, transcending the traditional boundaries of either discipline.

Religious experiences—whether mystical (the Unveiling) or philosophical (Kalam)—are often partial and subjective, rooted in the interpretation of the experiential or the choice of interpretation. Similarly, the dichotomy between pure philosophy and hard science has left gaps in our understanding, particularly when addressing the profound questions that defy disciplinary silos. What emerges is a need for a new mode of inquiry: one that integrates the precision of physics with the speculative breadth of philosophy, while remaining distinct from both.

This endeavor resides in an intermediate space, striving for a synthesis that we might call \textit{hard physical philosophy}\footnote{The term \textit{hard physical philosophy} draws an analogy to \textit{hard science fiction}, in which creative narratives are rooted in rigorous scientific facts. Similarly, \textit{hard physical philosophy} refers to philosophical narratives grounded firmly in hard physical facts.} or \textit{hard philosophical physics}\footnote{\textit{Hard philosophical physics} parallels \textit{hard philosophical fiction}, where creative narratives are deeply rooted in philosophical frameworks. In this context, \textit{hard philosophical physics} refers to physical theories or interpretations that are rooted in well-defined philosophical principles. This might be the same as philosophy of physics.}. It is a pursuit grounded firmly in theoretical physics yet oriented toward philosophical insight, aspiring to truths that transcend subjective belief and empirical constraints. As Jung expressed, “I know, I do not need to believe.” Through this balanced approach, the aim is to illuminate deeper truths about existence, forging a path that neither philosophy nor physics alone could fully explore.

\subsection{Introduction}
In this essay, we introduce Quantum Perspectivism—a novel interpretation that bridges the domains of quantum mechanics and Nietzschean perspectivism. It serves as both an interpretation of quantum mechanics, and a physical model for understanding Nietzsche’s profound ideas about perspectives and truth. By integrating quantum logic, complementarity, and contextuality with Nietzschean philosophy, Quantum Perspectivism redefines our understanding of observation, consciousness, and reality.

At the heart of this framework lies the Hilbert lattice, reimagined as Jung’s Unus Mundus—a unified ontological reality connecting observer and observed. Within this structure, perspectives and meta-perspectives are formalized as measurements, with each measurement corresponding to either a Boolean subalgebra (capturing classical, localized properties) or a Hilbert sub-lattice (representing meta-perspectives that synthesize multiple measurements into broader, quantum-like constructs). This distinction underscores the intricate interplay between localized (classical) perspectives and interconnected (quantum) meta-perspectives, revealing a layered structure of truth and understanding.

Nietzschean perspectivism posits that all truths are conditioned by cultural, historical, and psychological contexts, rejecting the extremes of relativism, where all perspectives are equally valid, and skepticism, which denies the possibility of knowledge. Similarly, Quantum Perspectivism navigates a middle ground between realism, which assumes pre-existing objective properties, and solipsism, which dismisses unmeasured observables (counterfactuals) as meaningless. In doing so, it highlights the complementarity and contextual dependencies that shape both philosophical and quantum truths.

The metaphor of perspectives as islands within the vast ocean of the Hilbert lattice, and meta-perspectives as continents synthesizing these islands, underscores the relational and constructed nature of knowledge. Just as Nietzschean perspectivism critiques the notion of an absolute, unchanging reality, quantum mechanics challenges classical objectivity, showing that perspectives are inherently partial and situated within a broader, interconnected framework.

This metaphor also emphasizes the ontological significance of perspectives in the Unus Mundus—a unified reality that encompasses all dualities. Here, perspectives and meta-perspectives not only serve as the foundation of the metaphysical observer’s understanding but also reveal that the Consciousness-World duality is a reflection of the metaphysical observer itself, or conversely, that the metaphysical observer emerges as a conceptual synthesis of this duality.

By uniting the reflective depth of philosophy with the empirical rigor of quantum mechanics, Quantum Perspectivism offers a dynamic interpretation of both fields. This interdisciplinary approach not only reveals the shared limitations of perspectivism in philosophy and physics but also provides a cohesive framework for exploring the dynamic relationship between observation, context, and reality.

\subsection{References}

The foundational concepts explored in this work draw upon a variety of sources that span philosophy, quantum mechanics, and their intersections. The \textit{Unus Mundus}, a unified ontological framework connecting all dualities such as psyche and matter, is inspired by Jung’s work on the nature of the psyche and his principle of \textit{synchronicity}, which emphasizes meaningful rather than causal connections \cite{Jung1971, Jung1973}. Nietzsche’s perspectivism, rejecting absolute truths in favor of perspectival, situated knowledge, provides the philosophical foundation for this model \cite{NietzscheGM, NietzscheWP}. Alan Schrift’s \textit{Nietzsche and the Question of Interpretation} offers a deeper exploration of Nietzschean perspectivism, emphasizing its interpretative and dynamic nature \cite{Schrift1990}.

The mathematical structure underlying quantum mechanics and quantum logic is deeply rooted in von Neumann and Birkhoff’s pioneering work on the logical foundations of quantum theory \cite{vonNeumann1936}, later formalized in von Neumann’s treatise \cite{vonNeumann1932} and expanded in comprehensive studies such as Svozil’s \cite{Svozil1998}. Mackey’s foundational contributions to quantum logic further elucidate its orthomodular structure and its role in quantum mechanics \cite{Mackey1963}.

The principle of complementarity, a cornerstone of quantum mechanics, was articulated by Bohr, who emphasized the necessity of combining mutually exclusive perspectives to achieve a holistic understanding of quantum phenomena \cite{Bohr1949}. In his earlier response to the EPR paradox, Bohr's seminal paper further clarifies the contextuality and complementarity inherent in quantum mechanics \cite{Bohr1935}. These ideas are foundational to the Copenhagen interpretation \cite{Bohr1935,Bohr1949}, systematically expounded in Paul Dirac’s \textit{The Principles of Quantum Mechanics} \cite{Dirac1958}, and rigorously analyzed in von Neumann’s \textit{Mathematical Foundations of Quantum Mechanics} \cite{vonNeumann1932}. Werner Heisenberg’s \textit{Physics and Philosophy} provides a philosophical synthesis of these principles, addressing the observer’s role and the interconnectedness of measurement and reality \cite{Heisenberg1958}.

The challenges to classical realism posed by quantum mechanics are encapsulated in the Einstein-Podolsky-Rosen (EPR) paradox \cite{Einstein1935} and addressed through Bell’s theorem, which rigorously demonstrates the incompatibility of local hidden variables with quantum predictions \cite{Bell1964}. Bell’s ideas, further elaborated in \textit{Speakable and Unspeakable in Quantum Mechanics} \cite{Bell2004}, and the Kochen-Specker theorem \cite{KochenSpecker1967}, underscore the contextuality and non-classical nature of quantum systems. N. David Mermin’s paper Hidden Variables and the Two Theorems of John Bell provides a lucid discussion of these results, making them accessible to both physicists and philosophers \cite{Mermin1993}. Asher Peres’s \textit{Quantum Theory: Concepts and Methods} further examines contextuality with detailed conceptual and mathematical treatments \cite{Peres1995}.

The interdisciplinary connection between psychology and quantum mechanics is enriched by the correspondence between Pauli and Jung, collected in Atom and Archetype: The Pauli/Jung Letters, which explores the Unus Mundus concept as a bridge between physics and metaphysics \cite{PauliJung1952}.

Thomas Nagel’s \textit{The View from Nowhere} offers a complementary philosophical perspective, addressing the limitations of objectivity and advocating for a perspective-based understanding of reality \cite{Nagel1986}.

Edwards, in his seminal work \cite{Edwards1979}, appears to have been the first to explore Quantum Perspectivism as a viable interpretation of quantum mechanics. His discussion provided a mathematical and philosophical foundation for understanding the relational nature of perspectives within quantum logic.

Quantum Perspectivism can be thought of as a middle ground between the Wigner-von Neumann interpretation \cite{vonNeumann1932,Wigner1961,Wigner1967,Wigner1962} and the Copenhagen interpretation \cite{Dirac1958,vonNeumann1932,Bohr1935,Bohr1949,Heisenberg1958}. While the former highlights consciousness as central to wavefunction collapse, and the latter emphasizes operational aspects of measurement, Quantum Perspectivism integrates elements of both, balancing ontological and epistemological perspectives. See \cite{Ydri:2020nys} for a related approach.

For a pedagogical introduction to these topics, including quantum logic, complementarity, and the philosophical implications of quantum mechanics, see the work, \textit{Philosophy and the Interpretation of Quantum Physics} \cite{Ydri2021}, and its synthesis \cite{Ydri:2021pjy}.

\subsection{Acknowledgments}
We acknowledge the assistance of GPT-4, an AI language model, in drafting this manuscript. The author retains full responsibility for all intellectual content, research, final edits, and creative decisions presented in this work. 

\section{A Brief Account of Nietzschean Perspectivism}

Nietzschean perspectivism is a profound and complex philosophical framework that challenges traditional notions of truth, reality, and knowledge. At its core, perspectivism asserts that there is no "view from nowhere," no absolute or objective standpoint from which reality can be fully grasped. Instead, all knowledge and understanding are conditioned by the perspective from which they arise, whether cultural, historical, and psychological.

Our goal is to reinterpret Nietzschean perspectivism through the lens of quantum mechanics, to provide a clearer and more accessible model for its principles. By drawing parallels between perspectivism and quantum concepts like contextuality, complementarity, and the relational nature of truth, we aim to illuminate its philosophical implications and explore its ontological foundations.  This effort not only bridges philosophy and physics but also offers a dynamic reinterpretation of perspectivism as a framework for understanding the interplay of perspectives in constructing reality.

Our goal is to reinterpret Nietzschean perspectivism through the lens of modern frameworks, such as quantum mechanics, to provide a clearer and more accessible model for its principles. By drawing parallels between perspectivism and quantum concepts like contextuality, complementarity, and the relational nature of truth, we aim to illuminate its philosophical implications and explore its ontological foundations. This effort not only bridges philosophy and physics but also offers a dynamic reinterpretation of perspectivism as a framework for understanding the interplay of perspectives in constructing reality. 

This reinterpretation will also serve as an interpretation of quantum mechanics, termed \textit{Quantum Perspectivism}, which highlights the perspectival nature of quantum truths and the contextual relationships inherent in quantum systems.


\subsection*{\underline{Core Ideas of Perspectivism}}

   \textbf{Truth as Perspective:}
   Nietzsche rejects the idea of objective, universal truth, asserting instead that all truths are perspectival:
   \begin{quote}
   "There are no facts, only interpretations." (\textit{The Will to Power} \cite{NietzscheWP}). 
   \end{quote}
   Truth is not an inherent property of the world but a construct shaped by the interpretative framework of the observer.

   \textbf{Multiplicity of Perspectives:}
   Reality cannot be fully comprehended from any single perspective. Instead, it requires a multiplicity of views, each offering a partial and situated understanding:
   \begin{quote}
   "The more affects we allow to speak about one thing, the more eyes, different eyes, we can use to observe one thing, the more complete will our ‘concept’ of this thing, our ‘objectivity,’ be." (\textit{On the Genealogy of Morality} \cite{NietzscheGM}).
   \end{quote}
   Nietzschean objectivity is not the absence of perspective but the inclusion of as many perspectives as possible.

   \textbf{Rejection of Absolute Being:}
   Nietzsche critiques metaphysical notions of a fixed, unchanging reality. Instead, he views reality as dynamic and ever-changing, revealed differently through various perspectives:
   \begin{quote}
   "There is no ‘being’ behind the doing, acting, becoming; the ‘doer’ is merely a fiction added to the deed—the deed is everything." (\textit{On the Genealogy of Morality}  \cite{NietzscheGM}).
   \end{quote}

   \textbf{Life-Affirming Perspectives:}
   Nietzsche emphasizes the importance of "life-affirming" perspectives—those that enhance vitality, creativity, and the will to power:
   \begin{quote}
   "To impose upon becoming the character of being—that is the supreme will to power." (\textit{The Will to Power} \cite{NietzscheWP}).
   \end{quote}

\subsection*{\underline{Challenges in Understanding Nietzschean Perspectivism}}

   \textbf{Ambiguity and Complexity:}
   Perspectivism resists reductive interpretations. It is neither relativism, which holds that all perspectives are equally valid, nor skepticism, which denies the possibility of knowledge.

   \textbf{Ontological Implications:}
   Does perspectivism imply that reality is just a sum of perspectives, or is there a deeper reality beyond perspectives?

   \textbf{Epistemological and Ethical Dimensions:}
   Perspectivism challenges traditional notions of knowledge and morality, making it simultaneously liberating and destabilizing.

   \subsection*{\underline{Summary}}
   
In summary, we have:

\begin{itemize}
    \item \textbf{Truth as Perspective:} All truths are perspectival, shaped by interpretative frameworks, rather than universal or objective facts.
    \item \textbf{Multiplicity of Perspectives:} Reality requires multiple, partial perspectives to form a fuller understanding.
    \item \textbf{Rejection of Absolute Being:} Reality is dynamic, characterized by becoming rather than a fixed state of being.
    \item \textbf{Life-Affirming Perspectives:} Not all perspectives are equal; those that enhance vitality and creativity are emphasized.
    \item \textbf{Ambiguity:} Perspectivism avoids the extremes of relativism and skepticism, demanding a nuanced interpretation.
    \item \textbf{Relational Nature of Knowledge:} Knowledge is relational, situated in cultural, psychological, and historical contexts.
\end{itemize}

\subsection*{\underline{Our Goals}}

\begin{itemize}
    \item \textbf{Provide a Physical Model:} Develop a quantum mechanical framework that parallels Nietzschean perspectivism, using quantum logic, contextuality, and complementarity to illustrate its principles. This model will also provide a very plausible interpretation of Quantum Mechanics.
    \item \textbf{Clarify Ontological Implications:} Reinterpret Nietzschean perspectivism in ontological terms, exploring how perspectives reveal or construct the nature of being.
    \item \textbf{Make Perspectivism Accessible:} By grounding perspectivism in a physical model, we aim to make its abstract ideas more comprehensible and applicable to contemporary thought.
      \item \textbf{Highlight Complementarity and Contextuality:} Demonstrate how perspectivism, like quantum mechanics, reveals the complementarity of seemingly incompatible facts and the contextual nature of truth and understanding.
\item \textbf{Explore Factual vs Counterfactual:} Examine the duality between factual and counterfactual truths, emphasizing how perspectivism integrates both into a coherent framework for understanding reality.

    \item \textbf{Unify Insights:} Draw connections between Nietzsche’s philosophy and modern scientific frameworks, offering a cohesive vision of perspectivism as a dynamic interplay of perspectives.
\end{itemize}

\section{Quantum Logic}

\subsection{A Brief Account}

Quantum logic provides a mathematical framework for understanding the logical structure of propositions in quantum mechanics. Unlike classical logic, which is governed by Boolean algebra, quantum logic arises from the structure of the Hilbert space and its lattice of subspaces. The first formulation of quantum logic was introduced by John von Neumann and Garrett Birkhoff in 1936, who proposed it as a non-classical logic to describe the relationships between quantum mechanical events \cite{vonNeumann1936}. This framework incorporates principles such as contextuality and complementarity, offering a richer and more nuanced understanding of measurement outcomes and their interdependence. In this section, we explore the key features of quantum logic, including its foundations, the role of Boolean subalgebras (blocks), and its philosophical implications for understanding the interplay between measurement and perspective.


\subsection*{\underline{Hilbert Lattice as Basis of Quantum Logic}}
Quantum logic arises from the structure of the Hilbert space, specifically its lattice of closed subspaces (the Hilbert lattice \( L \)). This lattice provides a framework for understanding the logical relationships between propositions about quantum systems.

\subsection*{\underline{Perspectives are Boolean Subalgebras (Blocks)}}
A block is a maximal Boolean subalgebra of the Hilbert lattice. It represents a specific measurement context, where all included observables are compatible (comeasurable). 
\begin{itemize}
    \item Each block corresponds to a single perspective on the system, governed by classical Boolean logic despite the probabilistic nature of quantum mechanics.
    \item The elements of a block are \textbf{atoms}---minimal propositions about the system, such as "spin-up along the \( x \)-axis."
\end{itemize}

\subsection*{\underline{Orthomodular Lattice Structure}}
The Hilbert lattice \( L \) satisfies the orthomodular condition, which generalizes the distributive property of classical Boolean logic. This makes quantum logic richer and more flexible than classical logic. 
\begin{itemize}
    \item Blocks within \( L \) are Boolean and locally measurable, but globally, they cannot be combined without considering quantum contextuality.
\end{itemize}

\subsection*{\underline{Pasting and Meta-Perspectives}}
A quantum non-Boolean logic is constructed by pasting together Boolean subalgebras (blocks). This process involves:
\begin{itemize}
    \item Identifying trivial and impossible propositions across blocks,
    \item Preserving the internal structure of each block.
\end{itemize}
The result is an orthomodular lattice that represents the combined quantum logic of all possible measurement contexts. These combined subalgebras correspond to \textbf{meta-perspectives}, offering a broader view of the quantum system.

\subsection*{\underline{Complementarity and Contextuality}}
\begin{itemize}
    \item Blocks representing measurements along different directions (e.g., \( x \), \( y \), \( z \)) are non-comeasurable but complementary in Bohr's sense. Together, they provide a richer understanding of the system.
    \item Contextuality, as demonstrated by the Kochen-Specker theorem, shows that the properties of the system depend on the specific measurement context. This implies that counterfactual observables (unmeasured properties) are not fully independent but are constrained by the measurement setup.
\end{itemize}

\subsection*{\underline{Limits of Perspectives}}
For simple systems like spin-\( \frac{1}{2} \), the combination of an infinite number of blocks can recover the entire Hilbert lattice (an \textbf{omni-perspective}). However, for more complex systems, even infinite measurements yield only partial information about the system, reflecting the inherent limitations of quantum knowledge.

Quantum logic captures thus the interplay between measurement, contextuality, and the structure of propositions in quantum mechanics. Boolean subalgebras (blocks) represent individual perspectives, while their combination through pasting defines the non-classical, orthomodular logic of quantum systems. The philosophical stance of quantum perspectivism emphasizes the relational nature of quantum measurements, balancing contextual constraints with conceptual utility.

\subsection*{\underline{Philosophical Implications}}
The relationship between measured and counterfactual observables can be understood through three philosophical perspectives:
\begin{itemize}
    \item \textbf{Realism:} Assumes no fundamental difference between measured and counterfactual observables, aligning with hidden variable theories.
    \item \textbf{Quantum Solipsism:} Views unmeasured observables as physically meaningless.
    \item \textbf{Quantum Perspectivism:} Acknowledges the mathematical and conceptual utility of counterfactual observables but emphasizes their relational constraints due to contextuality.
\end{itemize}
Quantum perspectivism provides a middle ground, advocating for a relational and contextual understanding of measurement outcomes. It balances the conceptual utility of counterfactual observables with the dependencies introduced by contextuality.


\subsection{A Physical Example: Spin-\(\frac{1}{2}\) Particle}
We start with a physical example.

Consider a spin-\( \frac{1}{2} \) particle as the observed system. An observer interacts with this system by measuring its spin along a specific direction, such as the \( x \)-axis. This particular measurement involves two compatible operators, \( S^2 \) and \( S_x \), which form the maximal set of comeasurable observables in this context. 

When the observer interacts with the system to measure the spin along the \(x\)-direction, they become entangled with it. In fact, they reduce the state of the system and observe either spin-up or spin-down, each with a computable probability.

This interaction defines, in a literal/linguistic sense, a \textbf{perspective} for the observer gained through a single measurement of the system.

Mathematically, this single perspective corresponds to a block of quantum logic.

Indeed, the above measurement corresponds to a 4-dimensional maximal Boolean subalgebra \( L_x \), referred to as a block and denoted by \( B \). Explicitly, this maximal subalgebra \( L_x \) corresponds to the set of all propositions about the spin along a specific axis (e.g., \(x\)):
\begin{itemize}
    \item \( P_+ \): The spin is up along the \(x\)-axis.
    \item \( P_- \): The spin is down along the \(x\)-axis.
    \item \( 1 \): The system is in some spin state (trivial proposition, always true).
    \item \( 0 \): The system is in no state (impossible proposition, always false).
\end{itemize}
These elements are minimal logical propositions, called \textbf{atoms}. They form a Boolean algebra, and no additional propositions about the spin along the \(x\)-axis can be added without violating the principles of Boolean logic.

This block is, in fact, a subalgebra of the so-called Hilbert lattice \( L \) of the Hilbert space \( H \). The Hilbert lattice \( L \) is the orthomodular lattice of the set \( C(H) \) of all closed linear subspaces of \( H \), which is isomorphic to the set \( P(H) \) of all projectors on \( H \). 

Quantum events correspond to experimental propositions \( p \) and are elements of \( C(H) \), denoted as \( M_p \), or equivalently are elements of \( P(H) \), denoted as \( P \). The subalgebra \( L_x \) is a subset of \( C(H) \) that is:
\begin{itemize}
    \item Closed under the logical operations: \( \wedge = \cap \), \( \vee = \oplus \), and \( \neg = \perp \),
    \item Contains the identity elements \( 1 \) and \( 0 \).
\end{itemize}
A block \( B \) of \( L \) is a \textbf{maximal Boolean subalgebra} of \( L \). Here, "maximal" indicates that the subalgebra contains the maximum number of atoms directly above \( 0 \), and "Boolean" signifies that the distributive law is satisfied.

The lattice \( L \) satisfies the orthomodular condition, which is a relaxation of the modular condition, itself a weaker form of the distributive law found in Boolean algebras. However, the block \( B = \{L_x\} \) is Boolean, and despite the inherent probabilistic nature of quantum mechanics, this measurement provides the observer with a single, classical-like perspective.

The observer has the free will to measure the spin along any other direction, such as \( y \) or \( z \), resulting in a different maximal Boolean algebra or block and, consequently, a new perspective on the same system. These blocks, representing different measurement directions and corresponding to distinct perspectives, are \textbf{non-comeasurable} but \textbf{complementary} in Bohr's sense.

More precisely, this complementarity implies that by combining Boolean subalgebras (blocks) \( L_{x_i} \)---pasting them together via their horizontal sum---we construct subalgebras of the Hilbert lattice that correspond to measurements of the spin of the particle along various directions \( x_i \). These subalgebras, which represent \textbf{meta-perspectives}, are explicitly given by the horizontal sum:
\begin{eqnarray}
    \mathbf{MO}_n = \bigoplus_{i=1}^n L_{x_i}.
\end{eqnarray}
This subalgebra is an orthomodular lattice and defines an instance of quantum logic. The set of all meta-perspectives is quantum logic.


A quantum non-Boolean logic is then constructed by pasting together Boolean subalgebras (blocks) \( L_{x_i} \) as follows:
\begin{itemize}
    \item The trivial propositions of all blocks are identified.
    \item The impossible propositions of all blocks are identified.
    \item Identical propositions in different blocks are identified.
    \item The logic and algebraic structures in all blocks remain the same.
\end{itemize}
In the specific case of a spin-\( \frac{1}{2} \) system, the limit of an infinite number of measurements corresponds to the entire Hilbert lattice, representing an \textbf{omni-perspective}. However, for more complex systems, even an infinite number of measurements will yield only a partial perspective, as the full state of the system remains inaccessible. This limitation is the typical case in quantum mechanics.

\subsection{Quantum Perspectivism}
Two key points are worth emphasizing. First, the reality hypothesis, which is challenged by Bell's theorem, asserts that the properties of the observed system have definite values prior to measurement. Indeed, measurements performed in experiments do not create their outcomes but only reveals pre-existing values.
 Second, the contextuality assumption, as demonstrated by the Kochen-Specker theorem, states that the properties of the system also depend on the specific "context" of other compatible (commuting) observables being measured. Quantum mechanics is inherently contextual because it does not adhere to the principles of realism.

In summary, every block is defined locally, i.e., it is locally measurable, but it cannot be defined globally without considering the simultaneous measurement of other blocks. At any given time, only one block can be measured from a collection of non-comeasurable blocks. This raises the question: is there a fundamental difference between an observable that is directly measured (the measured block) and one inferred counterfactually (from non-comeasurable blocks)?

Three philosophical positions address the relationship between measured and counterfactual observables:
\begin{itemize}
    \item \textbf{Realist Position (e.g., Einstein-Podolsky-Rosen):} There is no fundamental difference between measured observables and those inferred counterfactually, allowing all non-comeasurable blocks to be considered simultaneously.
    \item \textbf{Quantum Solipsism ("Unperformed Experiments Have No Results" \cite{Peres1995,peres1990}):} There is a strict fundamental difference, as unmeasured (counterfactual) observables are considered physically meaningless. In some sense, these counterfactual observables exist only in the mind and hence our characterization of this position as "solipsim". 
    \item \textbf{Quantum Perspectivism (Middle Ground):} This position acknowledges that counterfactual observables can make sense mathematically and conceptually but are bounded by contextuality. The measurement outcomes of one block depend on the context, including other blocks and their relationships. Thus, counterfactuality is not entirely free but is relationally constrained.
\end{itemize}
Quantum perspectivism balances the conceptual utility of counterfactual observables with the contextual dependencies revealed by Bell's theorem and the Kochen-Specker theorem. It avoids the extremes of both strict realism and solipsism, advocating for a relational and contextual understanding of measurement outcomes.



\subsection{Nietzschean Perspectivism as Quantum Perspectivism}

The analogy between quantum mechanics and Nietzschean perspectivism centers on the interplay between the observed system (the external World) and the observer (Consciousness). In quantum mechanics, the observed system, such as a spin-\( \frac{1}{2} \) particle, does not possess definite properties independently of measurement. Instead, the act of observation creates these properties by collapsing the system's state into a specific outcome. Similarly, in Nietzschean metaphysics, the World is not an objective reality independent of perception but is shaped and partially constructed by Consciousness through the act of experience/representation.

In the quantum framework, the observer and system become entangled during measurement, forming an inseparable relational entity. This entanglement mirrors the metaphysical entanglement between Consciousness and the World in perspectivism. Measurement in quantum mechanics reduces the state of the system, giving rise to a specific perspective—a classical logic that represents the observed property (e.g., spin-up or spin-down). Likewise, perception in Nietzschean terms reduces the state of the World into a particular perspective, comprising objective facts and subjective qualia.

Perspectives in quantum mechanics correspond to Boolean algebras (blocks), which represent classical views of the system. These perspectives are interlocked within a broader quantum logic, forming an orthomodular lattice. Similarly, perspectives in Nietzschean metaphysics are partial, context-dependent experiences/representations of the World, interwoven into a broader structure of interrelated experiences. 

Both frameworks acknowledge the impossibility of attaining a complete perspective: in quantum mechanics, this arises primarily from the failure of realism (the nature of physical reality) and secondarily from complementarity/contextuality (the nature of human knowledge); in perspectivism,  it stems from the relational and situated nature of consciousness and its understanding. As quantum mechanics is fundamentally concerned with the physical, this picture gives only a very plausible interpretation of quantum mechanics. In contrast, Nietzschean perspectivism extends further, addressing the nature of being itself by uniting external and internal realities into what Jung might call the Unus Mundus—a cohesive and unified reality encompassing all existence.

Finally, both frameworks emphasize the role of context. In quantum mechanics, measurement outcomes depend on the context of other compatible measurements. In Nietzschean perspectivism, the understanding of experiences is shaped not only by their psychological, sociological, historical, cultural, and situational context but also by how they relate to and are interpreted alongside other compatible experiences. Thus, both quantum mechanics and Nietzschean perspectivism highlight the constructed and relational nature of reality, rejecting absolute objectivity in favor of a dynamic interplay of perspectives.

\section{Hilbert Lattice as Jung's Unus Mundus and Persepctives as Islands}

\subsection{Perspectives as Islands in the Hilbert Lattice}
Let us consider first the simplified case of a spin-\( \frac{1}{2} \) particle.

Before measurement, the system exists in a quantum superposition of spin states, such as spin-up and spin-down along a given axis (e.g., \( x, y, z \)). By preparing to measure the spin along a particular axis, the observer defines the \textit{measurement context}, which determines the set of compatible observables to be measured. This context forms a \textit{perspective}, represented mathematically as a maximal Boolean subalgebra (block) within the quantum logic framework. In this perspective, the states of the system and observer are determined and mutually dependent, encapsulating the potential outcomes of the measurement.

A measurement causes the quantum state to collapse into one of the eigenstates of the measured observable. For instance, if the spin is measured along the \(x\)-axis, the state transitions into either spin-up or spin-down along \(x\). This collapse does not reveal a pre-existing property of the system but rather \textit{creates a property} that holds significance within the given perspective. The observer's choice of the measurement axis determines  which perspective is active and, consequently, which property is created.

The properties generated by the measurement are expressed within the logical structure of quantum logic through minimal logical propositions, forming a \textit{Boolean algebra}. This is a classical-like logical framework captures the relationships between the propositions about the system within the given perspective. The Boolean algebra exists as a localized structure within the broader Hilbert lattice, which encompasses all potential perspectives.

This framework has profound implications for understanding the nature of quantum measurements. It highlights the active role of measurement in defining the properties of the system. Perspectives are not neutral or absolute—they are \textit{relational constructs} that bind the observer and system states into a coherent framework. The mathematical description of perspectives within quantum logic emphasizes their \textit{contextual and contingent nature}, standing in contrast to the classical notion of objective, observer-independent properties. Instead of existing independently, properties emerge from the interplay of observer, system, and context, revealing the fundamentally perspectival nature of reality in the quantum domain.

The Boolean subalgebra, representing a specific perspective, is a subset of the broader quantum logic, structured as an orthomodular lattice. This lattice corresponds to the Hilbert space of the system and integrates all possible propositions about its state. While the Boolean subalgebra provides a self-consistent, classical-like logic within a given measurement context, the orthomodular lattice situates these individual perspectives within a unified framework, enabling their coexistence and interaction.

To visualize a perspective, imagine it as a localized "patch" or "region" or "island" within the larger Hilbert lattice. Each patch corresponds to a specific measurement context defined by the observer’s choice of compatible observables. Within this patch, the Boolean subalgebra governs the logical relationships between observed properties. Transitioning between patches reflects a shift in perspectives, governed by principles such as complementarity and contextuality. Complementarity ensures that incompatible perspectives, while exclusive, collectively enrich the understanding of the system. Contextuality emphasizes that properties observed in one perspective are shaped by the broader quantum logic, highlighting the interdependence of all perspectives within the Hilbert lattice. This visualization captures the perspectival and relational dynamics of quantum reality.


\subsection{Interpreting the Hilbert Lattice as the Unus Mundus}
The analogy between the Hilbert lattice and the Unus Mundus offers a profound conceptual framework for understanding the unity of perspectives and the interplay between observer and observed. The Hilbert lattice, representing the logical structure of quantum mechanics, can be mapped onto the Unus Mundus, envisioned by Jung as a unified underlying reality that connects all dualities, such as psyche and matter or observer and observed. In this interpretation, perspectives correspond to localized, contextual truths embedded within a broader ontological structure.

The Hilbert lattice, defined as the lattice of closed subspaces of a Hilbert space, represents all possible propositions about a quantum system. It serves as a comprehensive logical framework, integrating every potential measurement and perspective into a unified structure.

Within the Hilbert lattice, Boolean subalgebras correspond to specific perspectives, each defined by a maximal set of compatible (co-measurable) observables. These subalgebras provide self-consistent, classical-like snapshots of the system within the context of a particular measurement. They represent localized truths about the system and the observer’s interaction with it, acting as perspectival constructs that embed the observer and system within a specific context while remaining part of the larger unified structure.

Perspectives within the Hilbert lattice are interrelated through two key principles:
\begin{itemize}
\item Complementarity: Observables in one Boolean subalgebra are generally incompatible with those in another. These distinct perspectives, while mutually exclusive, collectively enrich the understanding of the system by emphasizing the necessity of multiple views.
\item Contextuality: The properties observed within a Boolean subalgebra are influenced by the broader framework of the Hilbert lattice, highlighting the relational nature of perspectives.
  \end{itemize}
Together, complementarity ensures that no single perspective can capture the entirety of reality, while contextuality underscores the interconnectedness of perspectives, each shaping and being shaped by the whole.

This concept is extensively discussed in Jung’s collaboration with Wolfgang Pauli in their exploration of the intersection between psychology and quantum physics.

A meta-perspective in the Hilbert lattice is constructed by integrating multiple perspectives, each corresponding to a Boolean subalgebra. This meta-perspective itself forms a Hilbert lattice, representing a broader but still incomplete view of the system, formed by combining complementary perspectives.

If individual perspectives are visualized as "patches" or "islands" within the vast ocean of the Hilbert lattice, meta-perspectives can be seen as "larger patches" or "continents," encapsulating the observer's accumulated measurements. However, these continents remain inherently smaller than the whole Hilbert lattice, which encompasses all possible perspectives. Even with the maximum conceivable number of measurements (theoretical infinity), the completed meta-perspective remains finite due to the constraints of complementarity and the contextual limitations of the observer.

The Hilbert lattice, as a whole, corresponds to Jung’s concept of the Unus Mundus, a unified ontological reality. This concept is extensively discussed in Jung’s collaboration with Wolfgang Pauli in their exploration of the intersection between psychology and quantum physics.

Hence, each meta-perspective, as a subset of the Hilbert lattice, represents an individual impression of the system as perceived by the observer’s consciousness. The collection of all such meta-perspectives associated with a single observer forms a construct that we identify as the metaphysical observer within the Unus Mundus. This metaphysical observer represents the totality of possible impressions of the system accessible to that observer’s consciousness.

This framework elucidates the role of  meta-perspectives as localized Hilbert lattices within the broader Unus Mundus. It emphasizes the interplay between the observed system, the conscious observer, and the inherent limitations of perspectives. By situating observers and systems within the perspectival ontology of the Unus Mundus, this model unifies the observed system and the conscious observer into a single metaphysical entity—the metaphysical observer—defined by the set of all meta-perspectives associated with the observer’s consciousness.


\subsection{Principle of Unus Mundus Relativity}

The Principle of Unus Mundus Relativity provides a metaphysical framework for understanding how perspectives and meta-perspectives coexist and interact within the unified ontology of the Unus Mundus, identified here as the Hilbert lattice. By drawing analogies with physical relativity, this principle explores how cognitive meta-perspectives serve as metaphysical "inertial" frames, ensuring consistent physical laws across observers, while qualia meta-perspectives introduce subjective, "non-inertial" dynamics. Together, these elements illuminate the interplay between objective and subjective realities, anchoring the relational unity of the Hilbert lattice as the Unus Mundus.

\subsection*{\underline{Metaphysical Inertial Reference Frames}}

In the Unus Mundus, each conscious observer generates a \textit{meta-perspective}, which serves as a localized Hilbert lattice—a patch of the larger Unus Mundus. These meta-perspectives represent the observer’s accumulated observations, measurements, and contextual understandings of the system. While distinct and specific to each observer, these meta-perspectives are interwoven within the Unus Mundus, reflecting the inherently relational nature of all perspectives.

Despite the individuality of meta-perspectives, cognitive representations (logical, rational, and conceptual frameworks) associated with different observers often yield equivalent descriptions of the physical laws and the world. This equivalence ensures coherence across distinct observers, anchoring a shared reality within the Unus Mundus. Cognitive meta-perspectives thus serve as \textit{metaphysical inertial reference frames}, providing a stable and consistent foundation for understanding reality across conscious observers.

By acting as metaphysical inertial frames, cognitive representations anchor reality by delivering consistent and equivalent descriptions of physical laws and the world. This mirrors the principle of relativity in physics, ensuring that reality, as perceived through cognitive meta-perspectives, appears invariant and coherent across all observers.

\subsection*{\underline{Principle of Relativity in the Unus Mundus}}

The \textit{Principle of Relativity in the Unus Mundus} asserts that physical laws and the structure of the world appear consistent across all metaphysical inertial reference frames, i.e., cognitive meta-perspectives associated with conscious observers. This principle parallels the physical principle of relativity, where the laws of physics remain invariant across all inertial frames. Cognitive meta-perspectives provide a metaphysical foundation for shared realities and universal physical laws, ensuring coherence within the relational fabric of the Unus Mundus across distinct observers.

\subsection*{\underline{Qualia Meta-Perspectives}}
In contrast, \textit{qualia meta-perspectives}, which capture subjective, phenomenal experiences, do not adhere to this principle. These are analogous to \textit{non-inertial reference frames} in physical relativity, where acceleration or external forces introduce distortions or variations in observed phenomena. Similarly, qualia meta-perspectives are unique to individual observers, resisting standardization or equivalence. Despite this non-equivalence, qualia meta-perspectives contribute a unique layer to the structure of reality, reflecting the subjective and individualized nature of phenomenal consciousness.

While cognitive meta-perspectives can be demonstrably equivalent across observers, the identity of qualia meta-perspectives cannot be empirically verified. At best, we postulate their equivalence to preserve the relational unity of the Unus Mundus. This postulation implies a universal structure underlying subjective experiences, even if their individual manifestations remain irreducibly distinct.

The interplay between cognitive meta-perspectives (metaphysical inertial frames) and qualia meta-perspectives (non-inertial frames) forms the relational unity of the Unus Mundus. This duality reflects the layered and interdependent nature of reality, balancing invariance and individuality. Cognitive meta-perspectives ensure shared understanding and coherence, while qualia meta-perspectives capture the unique, subjective aspects of experience.

By establishing equivalence across cognitive meta-perspectives and acknowledging the distinct role of qualia meta-perspectives, the \textit{Principle of Relativity in the Unus Mundus} provides a cohesive framework for understanding how distinct observers, perspectives, and representations coexist within the relational ontology of the Unus Mundus. This framework bridges the objective and subjective dimensions of reality, offering insights into the interplay between shared physical laws and the uniqueness of conscious experience, thereby reconciling quantum mechanics and perspectivism within a unified vision of existence.

\subsection{{Synchronicity}}

The perspectival ontology of the Unus Mundus incorporates Jung’s concept of \textit{synchronicity} to complement causality, rather than displace it. While causality governs the mechanical and physical interactions between observer and system, synchronicity emphasizes meaningful connections, particularly relevant in the mental and experiential domains.

In quantum mechanics, the collapse postulate provides a striking example of this dual framework. The measurement process not only involves causal interaction but also establishes a synchronistic relationship between the observer’s conscious state and the system’s observed property. This relationship is meaningful, as the act of observation actively defines the properties within the chosen perspective, reflecting the interplay between consciousness and the World.

Synchronicity also sheds light on the connection between mind and body, and by extension, between consciousness and the World. These realms are not solely bound by causality but resonate through synchronistic events, revealing a perspectival unity that integrates the observer’s experiences and interactions within the Unus Mundus. By bridging causality and synchronicity, this framework highlights the layered and multifaceted nature of the perspectival ontology.

\section{Concluding Remarks}

\subsection{Analogy: Quantum Perspectivism vs Nietzschean Perspectivism}
In summary, the following analogy can be drawn between quantum perspectivism and Nietzschean perspectivism (refer to table~\ref{table4} for details):

\begin{itemize}
    \item \textbf{Observed System and the World:} 
    \begin{itemize}
        \item In quantum mechanics, the system (e.g., a spin-\( \frac{1}{2} \) particle) represents the external World.
        \item In perspectivism, the World may or may not be fixed; however, as it is experienced, it is shaped by its interaction with Consciousness.

    \end{itemize}
    \item \textbf{Observer and Consciousness:}
    \begin{itemize}
        \item The observer in quantum mechanics corresponds to internal Consciousness in perspectivism.
        \item Both are entangled with their respective systems during observation or experience.
    \end{itemize}
  \item \textbf{Measurement and Perception:}
     \begin{itemize}
  \item Measurement in quantum mechanics collapses the system’s state, creating (not merely revealing) a specific set of properties described by a mathematical/physical perspective. This perspective combines objective facts, offering a coherent but partial understanding of the system. 

\item Perception in perspectivism reduces the state of the World, constructing (and perhaps creating) a particular set of subjective experiences (qualia) and cognitive representations described by a cognitive/psychological perspective. This perspective integrates objective facts and subjective qualia, providing a coherent yet incomplete understanding of the World. 
     \end{itemize}

    \item \textbf{Perspectives and Logic:}
    \begin{itemize}
        \item Perspectives in quantum mechanics are mathematical/physical perspectives given by Boolean subalgebras (blocks) interlocked into quantum logic (set of all orthomodular lattices).
        \item Perspectives in perspectivism are actual cognitive/psychological perspectives given by partial subjective experiences and cognitive representations interwoven into a broader structure of relational knowledge.
    \end{itemize}
  \item \textbf{Complementarity and Meta-Perspectives:}
    \begin{itemize}
    \item In quantum mechanics, a meta-perspective is formed by combining multiple measurements, reflecting the complementarity of different persecptives. However, it remains inherently partial, even with an infinite number of observations over time, as no single perspective can capture the entirety of the system.

    \item In perspectivism, a meta-perspective is constructed through the synthesis of experiences and representations, embodying the complementarity of diverse perspectives. Nevertheless, it remains fundamentally incomplete, even when considering an infinite amount of time or perceptions, as no single perspective or synthesis can encompass the whole of the World.
\end{itemize}

    \item \textbf{Context-Dependence:}
    \begin{itemize}
        \item Measurement outcomes depend on the context of other compatible measurements.
        \item Perceptions are influenced by cultural, psychological, and situational contexts. 
    \end{itemize}
  \item \textbf{Limits of Perspectivism:}
     \begin{itemize}
    \item Quantum mechanics asserts that a complete classical understanding of the system is unattainable; instead, only interwoven semi-classical perspectives and meta-perspectives can be formed.  
\item Perspectivism, analogously, asserts that a complete rational understanding of the World is unattainable; instead, only interwoven semi-rational perspectives and meta-perspectives can be formed.  
    \end{itemize}
\end{itemize}


\subsection{A Metaphysical Theory}

In summary,  the analogy between quantum perspectivism and Nietzschean perspectivism can be promoted into a metaphysical theory a follows:
\begin{itemize}
\item \textbf{Perspectives as Islands in the Hilbert Lattice}
\begin{itemize}
    \item Before measurement, the system exists in a superposition of states, and the observer defines the \textit{measurement context}, forming a perspective represented by a Boolean subalgebra.
    \item Measurement collapses the state into one of the eigenstates of the observable, creating a property meaningful within the chosen perspective.
    \item Perspectives are relational constructs, expressed as Boolean algebras, and are localized within the broader Hilbert lattice, governed by complementarity and contextuality.
\end{itemize}

\item \textbf{Hilbert Lattice as the Unus Mundus}
\begin{itemize}
    \item The Hilbert lattice represents all possible propositions about a quantum system and is analogous to the Unus Mundus, Jung's unified ontological reality.
    \item Boolean subalgebras correspond to localized perspectives, while meta-perspectives combine multiple subalgebras to form broader but still incomplete views of the system.
    \item Even with infinite measurements, a meta-perspective remains a finite patch within the infinite Hilbert lattice, reflecting complementarity and contextual constraints.
\end{itemize}

\item \textbf{Metaphysical Inertial Reference Frames}
\begin{itemize}
    \item Cognitive meta-perspectives act as \textit{metaphysical inertial frames}, ensuring consistent descriptions of physical laws across observers, akin to inertial frames in relativity.
    \item The \textit{Principle of Relativity in the Unus Mundus} asserts that physical laws and structures appear consistent across all metaphysical inertial frames.
\end{itemize}

\item \textbf{Qualia Meta-Perspectives and Non-Inertial Frames}
\begin{itemize}
    \item Qualia meta-perspectives, capturing subjective experiences, lack equivalence and resemble non-inertial frames in relativity.
    \item While their equivalence is unverifiable, postulating their unity preserves the relational coherence of the Unus Mundus.
\end{itemize}

\item \textbf{A Unified and Unifying Ontology of the Unus Mundus}
\begin{itemize}
    \item The interplay between cognitive meta-perspectives (inertial frames) and qualia meta-perspectives (non-inertial frames) establishes the relational/perspectival unity of the Unus Mundus.
    \item The \textit{Principle of Relativity in the Unus Mundus} bridges shared physical laws and individual consciousness, uniting objective and subjective dimensions of reality.
\end{itemize}
\item \textbf{Synchronicity}
\begin{itemize}
\item  \textit{Synchronicity}, as incorporated into the perspectival ontology of the Unus Mundus, complements causality by emphasizing meaningful connections, particularly in mental and experiential domains.
\item In quantum mechanics, the \textit{collapse postulate} exemplifies this dual framework, establishing a synchronistic relationship between the observer’s conscious state and the system’s observed property.
  \item This connection extends to the interplay between \textit{Mind and Body}, as well as \textit{Consciousness and the World}, revealing a perspectival unity that integrates experience and interaction within the Unus Mundus.
\end{itemize}
\end{itemize}

\begin{table}[H]
\centering
\begin{tabular}{|p{0.35\textwidth}|p{0.35\textwidth}|p{0.35\textwidth}|}
\hline
\textbf{Aspect}             & \textbf{Quantum Mechanics}                                    & \textbf{Nietzschean Perspectivism}                           \\ \hline
\textbf{Observed System and the World} 
& The system (e.g., a spin-\( \frac{1}{2} \) particle) does not possess definite properties until measured. The system is described by a quantum superposition, i.e. not fixed in the classical sense.
& The World may or may not be fixed; however, as it is experienced, it is shaped by its interaction with Consciousness. \\ \hline

\textbf{Observer and Consciousness}    
& The observer is entangled with the system during measurement, influencing and being influenced by the system's properties. 
& Consciousness is metaphysically entangled with the World during perception, shaping and being shaped by its interactions. \\ \hline

\textbf{Measurement and Perception}    
& Measurement collapses the system’s state, creating (not merely revealing) a specific set of properties described by a mathematical/physical perspective, offering a coherent but partial understanding of the system. 
& Perception reduces the state of the World, constructing subjective experiences and cognitive representations described by a cognitive/psychological perspective, providing a coherent yet incomplete understanding of the World. \\ \hline

\textbf{Perspectives and Logic}        
& Perspectives are represented by Boolean subalgebras (blocks) interlocked into quantum logic (orthomodular lattices), forming the mathematical framework of quantum mechanics. 
& Perspectives are represented by partial subjective experiences and cognitive representations, interwoven into a broader relational structure of knowledge. \\ \hline

\textbf{Complementarity and Meta-Perspectives}             
& A meta-perspective arises from combining multiple measurements but remains inherently partial, even with infinite observations (complementarity principle). 
& A meta-perspective emerges from synthesizing experiences and representations but remains fundamentally incomplete, even with infinite perceptions. \\ \hline

\textbf{Context-Dependence}            
& Measurement outcomes depend on the context of other compatible measurements. 
& Perceptions are shaped by cultural, psychological, and situational contexts, emphasizing the relational and situated nature of understanding. \\ \hline

\textbf{Limits of Perspectivism}       
& A complete classical understanding of the system is unattainable; only interwoven semi-classical perspectives and meta-perspectives can be formed. 
& A complete rational understanding of the World is unattainable; only interwoven semi-rational perspectives and meta-perspectives can be formed. \\ \hline
\end{tabular}
\caption{The Analogy Between Quantum Mechanics and Nietzschean Perspectivism.}
\label{table4}
\end{table}

\subsection{Conclusion}

Quantum Perspectivism can be thought of as occupying a middle ground between the Wigner-von Neumann interpretation \cite{vonNeumann1932,Wigner1961,Wigner1967,Wigner1962} and the Copenhagen interpretation \cite{Dirac1958,vonNeumann1932,Bohr1935,Bohr1949,Heisenberg1958}. See \cite{Ydri:2020nys} for a closely related approach.  While the Wigner-von Neumann interpretation emphasizes the ontological role of consciousness in the collapse of the wavefunction, the Copenhagen interpretation focuses on the epistemological (operational) aspects of quantum measurements, avoiding ontological commitments.

Indeed, perspectives in this model (associated with Boolean subalgebras for single measurements and Hilbert sub-lattices for meta-perspectives) formalize the act of measurement while emphasizing the synchronistic interrelation between observer and system in the Unus Mundus/Hilbert lattice. This duality ensures that Quantum Perspectivism does not rely exclusively on consciousness as the sole agent of collapse, as in the Wigner-von Neumann model, nor does it rely entirely to operationalism, as seen in the Copenhagen interpretation. Instead, it provides a dynamical framework that underscores both the ontological and epistemological dimensions of quantum mechanics.

\end{document}